# Single scan STEM-EMCD in 3-beam orientation using a quadruple aperture

Hasan Ali[1,2,¥,*], Sharath Kumar Manjeshwar Sathyanath[2], Cheuk-Wai Tai[1], Jan Rusz[3], Toni Uusimaki[1], Björgvin Hjörvarsson[3], Thomas Thersleff[1], Klaus Leifer[2]

[1]Department of Materials and Environmental Chemistry, Stockholm University, 106 91 Stockholm, Sweden

[2]Department of Materials Science and Engineering, Uppsala University, Box 534, 75121, Uppsala, Sweden

[3]Department of Physics and Astronomy, Uppsala University, Box 516, 751 20 Uppsala, Sweden

[¥]Present address: Ernst Ruska-Centre for Microscopy and Spectroscopy with Electrons and Peter Grünberg Institute, Forschungszentrum Jülich, 52425 Jülich, Germany

[*]Corresponding author: ali.hasan@angstrom.uu.se

The need to acquire multiple angle-resolved electron energy loss spectra (EELS) is one of the several critical challenges associated with electron magnetic circular dichroism (EMCD) experiments. If the experiments are performed by scanning a nanometer to atomic-sized electron probe on a specific region of a sample, the precision of the local magnetic information extracted from such data highly depends on the accuracy of the spatial registration between multiple scans. For an EMCD experiment in a 3-beam orientation, this means that the same specimen area must be scanned four times while keeping all the experimental conditions same. This is a non-trivial task as there is a high chance of morphological and chemical modification as well as non-systematic local orientation variations of the crystal between the different scans due to beam damage, contamination and spatial drift. In this work, we employ a custom-made quadruple aperture to acquire the four EELS spectra needed for the EMCD analysis in a single electron beam scan, thus removing the above-mentioned complexities. We demonstrate a quantitative EMCD result for a beam convergence angle corresponding to sub-nm probe size and compare the EMCD results for different detector geometries.

## 1. Introduction:

The discovery and subsequent experimental realization of electron magnetic circular dichroism (EMCD) measurements [1] in the transmission electron microscope (TEM) opened the door to characterize the local structure and magnetic properties of magnetic materials at nanometer to atomic scale. The experimental development over the last more than 15 years [2] made it possible to characterize the magnetic moments of single atomic planes [3] [4] and atomic site-specific magnetic measurements [5]. The technique has been used to detect the magnetic ordering in an antiferromagnet [6] and has been used to correlate the magnetic properties to structural changes at nano scale [7] [8] [9] [10] [11]. Despite the remarkable progress in optimizing and improving the experimental conditions [12] [13] [14] [15] [16] [17], there are still critical challenges associated with the EMCD experiments which enhance the level of difficulty in carrying out and interpreting the results of such experiments. The major challenges include poor signal to noise ratio of the EMCD signal, its strong dependence on dynamical diffraction effects, high sensitivity to crystal orientation conditions and multiple acquisitions needed in an EMCD experiment.

In the initial experimental setup proposed by Schattschneider et al. [1], TEM specimen is tilted to a 2-beam condition (2BC) and subsequently two electron energy loss spectra (EELS) are acquired at conjugate scattering angles. The EMCD signal is obtained by taking the difference of these two EELS spectra after standard post-processing. This is perhaps the simplest and experimentally most convenient diffraction condition as it can yield maximum EMCD signal strength and needs only two acquisitions. However, there are certain critical challenges associated with this condition which arise due to the asymmetry of the 2BC [18] [19]. This asymmetry leads to strong dependence of the EMCD signal strength on slight changes in the diffraction conditions. In our previous work using a customized aperture, we showed that a misalignment as small as 2 mrad from the perfect 2BC is sufficient to significantly diminish the EMCD signal [20]. Due to slight imperfections in crystal structures, there are high chances to encounter such tiny mistilts specially in a STEM map. In addition to weakening the strength of the EMCD signal, these small orientation changes may also lead to difficulties in post-edge normalization (discussed below), making the quantitative analysis cumbersome. So perhaps this diffraction geometry is suitable mainly for the detection of EMCD signal or the qualitative analysis but poses critical challenges for a quantitative analysis.

EMCD experiments can also be carried out under zone axis conditions which ensures that the atomic columns are parallel to the electron beam, opening the prospects for atomically resolved EMCD measurements, which are not possible under 2 or 3-beam conditions. However, executing EMCD experiments under a zone axis is very challenging due to the complex distribution of the magnetic signal around the Bragg spots [21] [22]. In addition, the relative strength of the magnetic signal is much lower compared to 2 or 3-beam conditions. In our previous work, we used complex shaped hardware apertures to detect the zone axis EMCD signals [23] [24]. The EMCD signal in a zone axis is also strongly influenced by small orientation changes, shown theoretically [19] and experimentally [23]. Due to the severe challenges associated with zone axis EMCD, this approach is the most challenging from an experimental point of view, ideally reserved for scientific questions where atomic resolution is truly required.

Another possible diffraction geometry for the EMCD experiments is a 3-beam condition (3BC) where two diffracted spots ± **G** are equally excited around the direct beam. The EMCD signal has four magnetic components in this case marked by white circles in **Fig 1 (c)**. Here ++ and -- represent the EELS spectra carrying positive chirality whereas +- and -+ represent the EELS spectra with negative chirality. In a conventional 2BC, the EMCD signal is obtained by taking a single difference between either ++ and +- or between -+ and -- whereas in this case, the two differences are added together to obtain a double difference. It has been shown that the double difference approach is more robust against the small misalignments of the crystal [18] and largely mitigates the asymmetric effects encountered in single difference case [19]. Although the magnetic signal strength is comparatively weaker in the 3BC than the 2BC [12], it is nevertheless sufficient for quantitative EMCD analysis. Here, the robustness against the asymmetric effects makes it the most favorable overall choice for quantitative EMCD experiments.

Another important factor to consider is the obtainable spatial resolution of analysis which makes EMCD superior to complementary techniques such as XMCD [25]. The initial experimental setup proposed by Schattschneider et al. [1] uses a parallel electron beam to illuminate the sample. The spatial resolution of EMCD analysis is defined by the diameter of the electron beam which, in the case of parallel illumination, is limited to few tens of nanometers or at the best can reach to a couple of nanometers by converging the beam [26]. The alternative route to achieve higher spatial resolution is to carry out the EMCD experiments in scanning TEM (STEM) mode. In STEM-EMCD experiments [27], a finely focused electron probe is scanned across a well-defined area of the specimen and an EELS spectrum is acquired at each probe position, for multiple off-axis detector positions. With the recent advances in probe-correctors, it is possible to focus the probe to be smaller than low order atomic plane distances, permitting atomic-plane EMCD measurements to be performed when the sample is tilted to 2BC or 3BC [4] [28]. In addition to improving the spatial resolution, STEM-EMCD allows for advanced statistical methods to be employed by acquiring hundreds of EELS spectra compared to single measurements done in the TEM mode, a greatly improved signal to noise ratio by integrating multiple spectra and an efficient dose distribution on the analyzed area of the sample [7] [29].

One highly problematic part in such an experiment is that the same area of the specimen must be scanned multiple times to acquire multiple STEM-EELS datasets at different off-axis aperture positions. The exact number of scans depends on the diffraction condition chosen for the experiment. To correlate the spectra in the multiple STEM-EELS datasets on a pixel-by-pixel basis, all the experimental conditions must remain the same during their acquisition, which is a non-trivial task. A slight specimen drift in between the multiple sequential scans can degrade the spatial registration and/or the orientation conditions under the electron probe. Also, the morphology of the specimen can change between multiple scans e.g. due to damage or contamination caused by high intensity electron probe. An ideal situation to remove all these complexities would be to acquire all the EELS spectra needed in an EMCD experiment in a single beam scan. While we have previously demonstrated such single pass STEM-EMCD experiments for the 2BC [30] [20] and zone-axis conditions [24] [23], the quantitative EMCD experiments under these conditions are challenging due to reasons discussed above.

Considering the advantages of double difference procedure, we here extend our experience to model single pass STEM-EMCD experiments for 3-beam diffraction geometry. We report an experimental setup using a quadruple aperture to simultaneously acquire four angle-resolved STEM-EELS datasets in a single scan. The application of double difference procedure on these four datasets yields a quantitative EMCD signal. We show that the double difference procedure to obtain the EMCD signal is more tolerant to small orientation changes and reduces complexities in the post-processing of the EELS spectra.

## 2. Experimental Details:

We used a 35 nm bcc Fe thin film grown on MgO (001) substrate for the investigations. The sample was fabricated as described in [24]. The TEM specimen was prepared in a plan view geometry by mechanical polishing, dimple grinding and subsequent Ar-ion milling to reach perforation. The EMCD experiments were performed on a JEOL 2100F equipped with a post column Gatan Tridiem imaging filter (GIF). The microscope was operated at an acceleration voltage of 200 kV. The quadruple aperture was installed at the spectrometer entrance plane in such an orientation that the four aperture holes do not overlap along $q_y$-axis in the diffraction plane. The microscope was switched to STEM mode and the convergence semi-angle was set to 7.5 mrad by changing the current in mini-condenser lens. The TEM specimen was tilted to a 3-beam condition from the [001] zone axis and manually rotated in the holder to orient the Bragg discs with respect to the four aperture holes according to the EMCD experimental requirements as shown in **Fig 1(c)**. A 120 x 120 $nm^2$ well-oriented region of the specimen was chosen and the electron probe was scanned across this region. An in-house developed script was used to acquire the GIF CCD image at each probe position while keeping the GIF in EELS mode. This acquisition results in a 4DSTEM dataset where the first two dimensions represent the spatial coordinates of the specimen area whereas the later two dimensions are the energy loss and the momentum transfer along $q_y$ as shown in **Fig 2**.

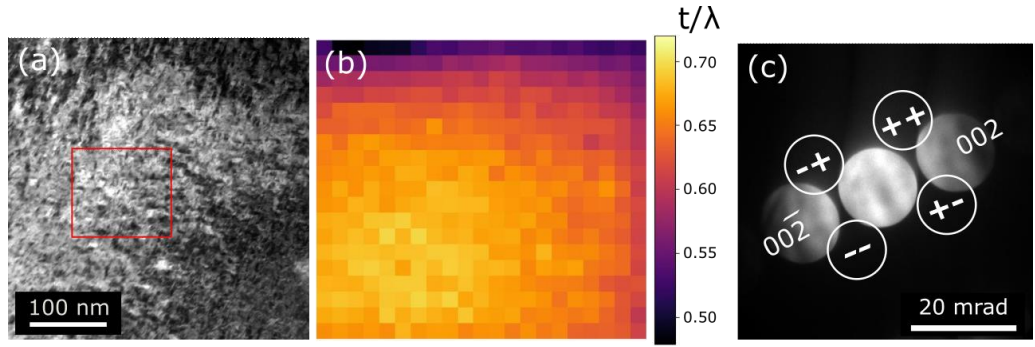

*Fig 1. Experimental Conditions for the single scan STEM-EMCD experiment. (a) survey image showing the area scanned during the experiment in a red box (b) thickness in terms of t/λ for the sample area used in the experiment (c) diffraction conditions showing the 3-beam orientation with g = ±002 excited. The aperture positions for the EMCD experiment are shown by white circles. The quadruple aperture was oriented such that the four aperture holes overlay these positions.*

More information about such data acquisition mode can be found at [20] [24] [31] [30]. The same area of the specimen was scanned once more to acquire the low-loss EELS spectra which were used for thickness determination. The low-loss EELS data was acquired on-axis by inserting standard circular aperture of the spectrometer.

## 3. Results:

**Fig 2** shows the CCD image of the quadruple aperture overlaid on the diffraction pattern (similar to **Fig 1(c)**). The four aperture holes are named as A1, A2, A3 and A4. Switching to EELS mode and acquiring the CCD image under these conditions produces a 2D EELS image where the $q_x$ axis is replaced by the energy dispersive axis and $q_y$ axis, with a scaling factor, is preserved as shown in **Fig 2**. The information along $q_x$ is integrated and cannot be resolved. In the resulting 2D EELS image, four distinct $q_y$ resolved intensity bands (marked by ++, +-, -+, --) corresponding to the four aperture holes can be observed. There is a slight overlap between these bands as the aperture holes were slightly overlapping along $q_y$. The four EELS spectra needed for the EMCD analysis were extracted from the 2D EELS image by integrating the intensity of each band along the energy loss axis, avoiding the overlapping regions. The four extracted spectra are also shown in **Fig 2**.

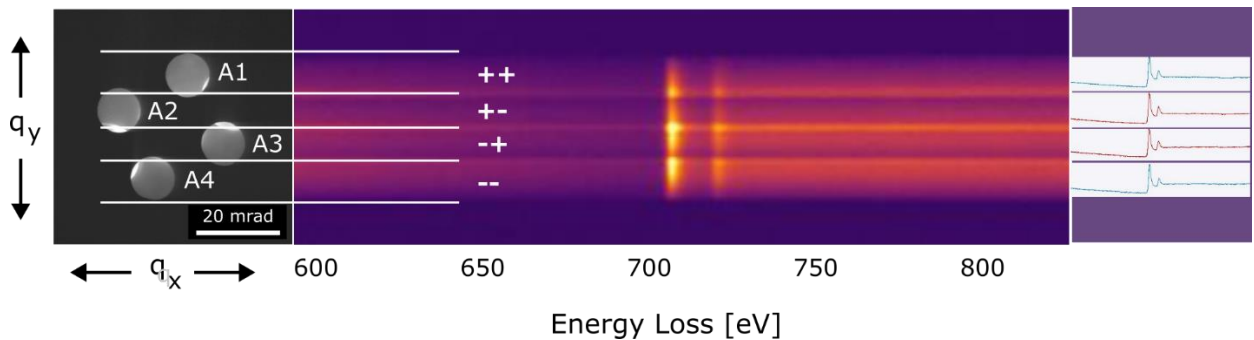

*Fig 2. Left: CCD image of quadruple aperture overlaid on the diffraction pattern, middle: 2D EELS image obtained by integrating the data of all the pixels in the STEM-EMCD map, such an image was acquired at each position while scanning the probe on the sample, right: momentum-resolved raw EELS spectra extracted from the 2D EELS image.*

The relative thickness of the sample was determined in terms of t/λ from the low-loss EELS dataset acquired from the same region where EMCD experiment was performed. The relative thickness map is shown in **Fig 1(b)**. It can be noted that the sample is much thicker than the optimum thicknesses of Fe for EMCD measurements [18]. Unfortunately, we found only this area to be well-oriented for the EMCD experiment after setting up all other experimental conditions such as rotating the sample, using free-lens control to set the convergence angles and aligning STEM for these settings. Nevertheless, the thickness of the magnetic Fe film was 35 nm everywhere and the rest of the thickness is contributed by the MgO substrate and we deconvolved the core loss EELS spectra to remove the plural scattering effects. Another interesting feature in the thickness map is the increasing thickness gradient from top to bottom although from our observations during the experiment, the thickness of the mapped area should be more or less homogeneous. As the low-loss EELS spectra were acquired after the EMCD map, a part of this thickness gradient is likely the result of contamination built on the sample during the first scan. This highlights one of the advantages of a single-pass EMCD experiment as the multiple scans needed for conventional STEM-EMCD will see a systematic contamination build-up following each individual scan, complicating the subsequent analysis (particularly regarding the removal of plural scattering) and diluting the already weak EMCD signal.

**Fig 3** shows the four raw EELS spectra extracted from the 2D EELS image shown in **Fig 2**. The corresponding deconvolved spectra are also shown in the same figure. It can be seen that the intensity of the post-edge after Fe-$L_{2,3}$ edges goes significantly down in the deconvolved spectra indicating the efficient removal of plural scattering effects. Another observation is that the number of electron counts in second and third spectra is higher than the first and fourth spectra.

This is a consequence of slightly misaligned diffraction pattern with respect to the aperture holes which results in higher intensities within the diffracted discs' contributing more to aperture A2 and A3 than the other two apertures as can be seen in **Fig 2**. Nevertheless, this difference is arising mostly from the non-magnetic signal and can be removed by normalizing the post-edge intensity of the spectra after background subtraction. As mentioned above, we expect the double difference procedure to largely eliminate such effects caused by asymmetry of the diffraction geometry as well as detectors' position.

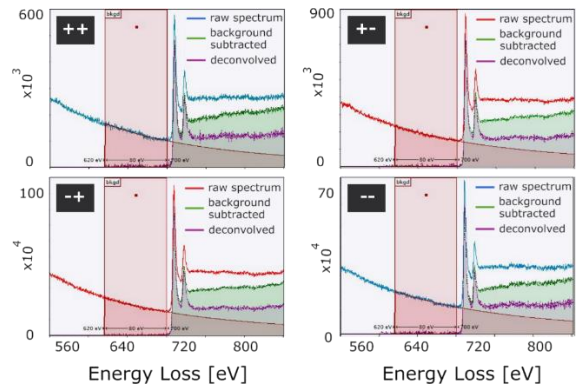

*Fig 3. Raw, background subtracted and deconvolved EELS spectra extracted from the integrated STEM-EMCD map for the four aperture positions.*

To test whether the double difference procedure really helps to compensate the asymmetry and misorientation effects and results in a better EMCD signal than the single difference, we divided the data into three categories. For the first two cases, we extracted the EMCD signals by taking the single differences between aperture positions A1, A3 and A4, A2 respectively as used in the conventional 2BC EMCD experiments. In the third case, we extracted the EMCD signal by the double difference of all the four apertures. **Fig 4** shows the maximum EMCD signal seen at $L_3$ edge for each pixel in the map in these three cases. There is a reasonable EMCD signal in case (a) whereas it gets very weak and disappears at most of the pixels for case (b). These significant differences between the two maps are caused by slight changes in underlying diffraction conditions. In this case, the crystal was probably misoriented from the 3-beam condition in a way that $g = 002$ was closer to 2BC, improving the EMCD signal found at aperture positions A1-A3 whereas the EMCD signal at aperture positions A4-A2 went very weak as $\bar{g} = 00\bar{2}$ was far away from 2BC. When doing the STEM-EMCD experiment in a 2-beam condition, there is a decent probability to fall into situation (b) and end up with a very weak or no EMCD signal. However, the map obtained by double difference (**Fig 3(c)**) shows the best results as not only the EMCD signal is recovered at most of the pixels, but the strength of the signal also goes up compared to the other two maps. This means the double difference eliminates to a great extent, the negative effects induced by orientation changes or asymmetry in agreement with simulations and experiment [18] [19].

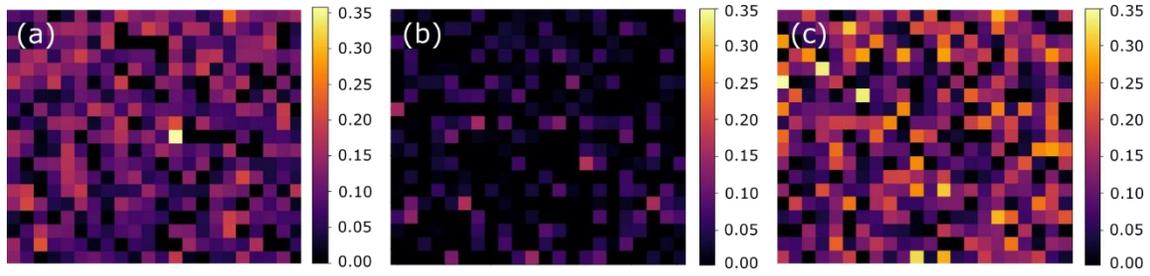

*Fig 4. Maximum EMCD signal seen at Fe-$L_3$ edge for single differences between (a) A1-A3 (b) A4-A2 and (c) the double difference.*

Another problem in the processing of EELS spectra to obtain the EMCD signal is encountered while normalizing the post-edge intensity of two chiral spectra, under the assumption that the post-edge does not contain magnetic information. For a reliable EMCD signal, the post-edges of the two spectra must be very well aligned after the normalization, as any differences in the post-edge slope will result in a residual post-edge slope in the EMCD signal, complicating the quantification. In our experience with the EMCD experiments, the post-edge of the two spectra is often not perfectly aligned after normalization. Out of curiosity, we checked the post-edge normalization for the three cases mentioned above. We used a normalization window of 50 eV in the interval 750-800 eV. We observed that after the normalization, the post-edge does not align very well for the single difference cases (between A1-A3 and A4-A2 apertures) whereas the post-edge aligns very well, even up to an extended energy range when the double difference procedure is applied as shown in **Fig 5**. This further emphasizes the advantage of double difference procedure for quantitative EMCD.

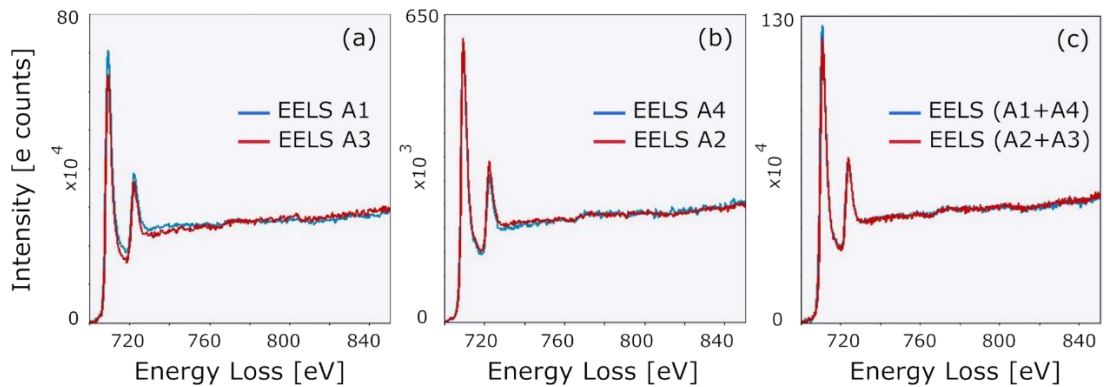

*Fig 5. Post-edge normalized EELS spectra for single differences (a) A1-A3 (b) A4-A2 and (c) the double difference procedure*

Finally, we extracted the overall EMCD signal by integrating all the data points in the deconvolved STEM-EMCD map and applying the double difference procedure. The result is shown **Fig 6**. The curve fitting to the EMCD signal and magnetic orbital to spin moments' ratio ($m_L/m_S$) calculation was done by the MATLAB code described in [23]. The resulting value of $m_L/m_S$ = 0.07 for bcc Fe is close to the values reported by previous EMCD experiments [12] [20] [11] and XMCD [32].

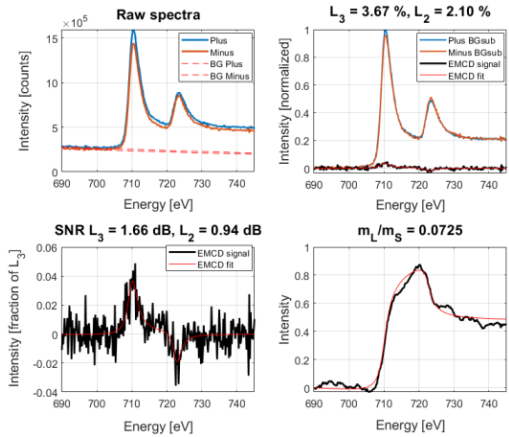

*Fig 6. EMCD signal extracted from the STEM-EMCD dataset after applying deconvolution and double difference method*

## 4. Discussion:

In our previous EMCD work under a 2-beam condition [20], we have shown how strongly the EMCD signal varies as a function of small changes in crystal orientations. The strong differences seen between the EMCD signals at aperture positions A1-A3 and A4-A2 are most probably a result of such slight misorientations in the region scanned by the electron beam. Considering that most of the crystals are not perfect single crystals, such small spatial deviations would be encountered in almost every experiment. The misalignment of post-edge regions due to different slopes is also a contributing factor to the strong variations seen in single difference maps. This post-edge misalignment may be a consequence of orientation changes but is not confirmed at the moment and needs to be explored in more detail. The point to be noted is that taking the double difference in the same data results in a more reliable EMCD signal and a very well aligned post-edge, simplifying the quantification process. So, we reiterate the worth of double difference approach for quantitative EMCD analysis.

The EMCD signal shown in **Fig 6** is still quite noisy. Apart from the inherently weak signal to noise ratio of the EMCD signal, there are several other reasons for the noise. The convergence semi-angle used for this experiment is relatively large (7.5 mrad) and the EMCD signal strength goes down with increasing convergence angles [27]. Moreover, the experiments were performed on a JEOL 2100 F equipped with a Gatan tridiem energy filter. This microscope does not support a nanobeam diffraction mode and the convergence angles resulting in normal STEM mode were too large for the EMCD experiment. To lower the convergence angles, we tuned the mini-condenser lens setting in combination with other condenser lenses and used the smallest condenser aperture. Despite of these optimizations, we could not go down more than 7.5 mrad as further changing the current in minicondenser lens caused scan distortions. After these manual settings, we ended up with a probe current of only few picoamps which is not optimum for the EMCD experiment. We compensated the low probe current by increasing the dwell time and used 1 s per pixel for the EMCD acquisition. Additional difficulties were encountered due to the unavailability of magnification reduction function while using GIF. Even the lowest camera length available in the microscope was too large on GIF CCD for the EMCD experiment. The camera length was significantly lowered by changing the currents in intermediate and projector lenses. The sample was manually rotated in the TEM sample holder to align the diffraction pattern in a pre-determined orientation before the start of other experimental settings. All these experimental challenges contributed to reduction of the EMCD signal seen at the end. Despite these difficulties, we could clearly demonstrate a double difference signal being more stable, giving a better signal to noise ratio as well as $m_L/m_S$ ratio in agreement with previous publications. Thus, the same aperture setup if implemented to a probe corrected system should be able to acquire EMCD signals with much higher quality and with convergence angles resulting in atomic resolution.

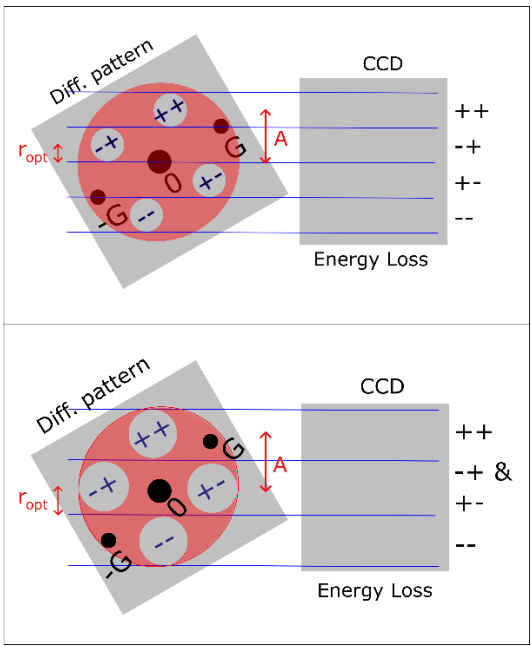

*Fig 7. Up: the quadruple aperture geometry used in this experiment. Down: an alternative aperture geometry allowing larger collection angle of the apertures.*

Another factor influencing the signal to noise ratio of EMCD signal is the collection angle of each aperture hole. If we consider **A** as the distance between the center of direct beam to the center of detector (++), the optimum radius for each collection aperture is limited to $r_{opt} = A/3$, under the conditions that the apertures do not overlap to each other as shown in **Fig 7**(up). In this experiment, we wanted to verify that the double difference procedure suppresses the asymmetry artefacts and is efficient for a clean EMCD signal extraction, so it was desired to have four individual spectra for this work. Once we establish that the method works well, one could use an alternative 4-hole design where only three individual spectra are extracted from CCD while still realizing the double difference extraction of EMCD signal (**Fig 7**(down)). In this geometry, we integrate the -+ and +- signals which have the same chirality. The optimum radius of each aperture allowed by this geometry is $r_{opt} = A/2$ which is 1.5 times more than the maximum radius allowed by the geometry used in this experiment. This means an overall improvement of 2.25 times in collection efficiency. Thus, this alternative setup would offer higher signal counts, resulting in an improved signal to noise ratio.

## 5. Conclusions:

We have reported an experimental setup to complete the EMCD experiment is a single STEM scan under 3-beam orientation conditions. With the help of a quadruple aperture, we simultaneously acquire the four momentum-resolved EELS spectra required for the EMCD signal extraction. This can greatly help to remove the EMCD experimental complexities and allow us to precisely correlate the local information contained in the four EELS datasets. We quantify the EMCD signal acquired for bcc Fe with a beam convergence angle of 7.5 mrad. The setup should be capable to acquire the EMCD signals with higher convergence angles reaching to atomic plane resolution [28].

## Acknowledgements:

H.A, C-W.T and T.T acknowledge the financial support from Swedish Foundation for Strategic Research SSF (ITM17-0301). H.A. gratefully acknowledges Swedish research council (project nr. 2021-06748) for their support. T.T acknowledges funding from the Swedish Research Council (project nr. 2016-05113). J.R. acknowledges Swedish Research Council (project no.2021-03848), Carl Tryggers Foundation and STINT for financial support.